\documentclass[10pt, conference, compsocconf]{IEEEtran}

\usepackage{amsmath}
\usepackage{program}
\usepackage{graphicx}
\usepackage[english]{babel}
\usepackage{algorithmic}
\usepackage{algorithm}
\usepackage{subfigure}

\usepackage[dvips]{color} 
\definecolor{blue}{rgb}{0,0,1}
\definecolor{green}{rgb}{0,1,0}
\definecolor{black}{cmyk}{0,0,0,1}

\begin{document}
\title{Parallelization Strategies for Ant Colony Optimisation on GPUs}

\author{\IEEEauthorblockN{Jos\'e M. Cecilia, Jos\'e M. Garc\'ia}
\IEEEauthorblockA{Computer Architecture \\ Department\\
University of Murcia\\
30100 Murcia, Spain\\
Email: \{chema, jmgarcia\}@ditec.um.es}
\and
\IEEEauthorblockN{Manuel Ujald\'on}
\IEEEauthorblockA{Computer Architecture\\ Department\\
University of Malaga\\
29071 M\'alaga, Spain\\
Email: ujaldon@uma.es}
\and
 \IEEEauthorblockN{Andy Nisbet, Martyn Amos}
\IEEEauthorblockA{Novel Computation Group\\
Division of Computing and IS \\
Manchester Metropolitan University \\
Manchester M1 5GD, UK\\
Email: \{a.nisbet,m.amos\}@mmu.ac.uk}}

\maketitle

\begin{abstract}

Ant Colony Optimisation (ACO) is an effective population-based meta-heuristic for the solution of a wide variety of problems. As a population-based algorithm, its computation is intrinsically massively parallel, and it is therefore theoretically well-suited for implementation on Graphics Processing Units (GPUs).  The ACO algorithm comprises two main stages: \textit{Tour construction} and \textit{Pheromone update}. The former has been previously implemented on the GPU, using a task-based parallelism approach. However, up until now, the latter has always been implemented on the CPU. In this paper, we discuss several parallelisation strategies for {\it both} stages of the ACO algorithm on the GPU. We propose an alternative {\it data-based} parallelism scheme for \textit{Tour construction}, which fits better on the GPU architecture. We also describe novel GPU programming strategies for the \textit{Pheromone update} stage.  Our results show a total speed-up exceeding 28x for the \textit{Tour construction} stage, and 20x for \textit{Pheromone update}, and suggest that ACO is a potentially fruitful area for future research in the GPU domain.
\end{abstract}

\section{Introduction}

Ant Colony Optimisation (ACO) \cite{ACOBOOK} is a population-based search method inspired by the behaviour of real ants. It may be applied to a wide range of hard problems \cite{dorigo2006ant,blum2005ant}, many of which are graph-theoretic in nature. It was first applied to the Travelling Salesman Problem (TSP) \cite{lawler1987traveling} by Dorigo and colleagues, in 1991  \cite{aco-tr,DORIGO2}. 

In essence, simulated ants construct solutions to the TSP in the form of {\it tours}. The artificial ants are simple agents which construct tours in a parallel, probabilistic fashion. They are guided in this by simulated {\it pheromone trails} and {\it heuristic information}. Pheromone trails are a fundamental component of the algorithm, since they facilitate indirect communication between agents via their {\it environment}, a process known as {\it stigmergy} \cite{dorigoa2000ant}. A detailed discussion of ant colony optimization and stigmergy is beyond the scope of this paper, but the reader is directed to \cite{ACOBOOK} for a comprehensive overview.

ACO algorithms are population-based, in that a collection of agents ``collaborates" to find an optimal (or even satisfactory) solution. Such approaches are naturally suited to parallel processing, but their success strongly depends on both the nature of the particular problem and the underlying hardware available. Several parallelisation strategies have been proposed for the ACO algorithm, on both shared and distributed memory architectures \cite{STUZLE98,JUNYONG,LIN}. 

The {\it Graphics Processing Unit} (GPU) is  a major current theme of interest in the field of high performance computing, as it offers a new parallel programming paradigm, called \textit{Single Instruction Multiple Thread} (SIMT) \cite{Tesla}. The SIMT model manages and executes hundreds of threads by mixing several traditional parallel programming approaches. Of particular interest to us are attempts to parallelise the ACO algorithm on the Graphics Processing Unit (GPU) \cite{ACOGPU1,ACOGPU2,ACOGPU3}.  These approaches focus on accelerating the {\it tour construction} step performed by each ant by taking a {\it task-based} parallelism approach, with pheromone deposition calculated on the CPU. 

In this paper, we fully develop the ACO algorithm for the Travelling Salesman Problem (TSP) on GPUs, so that {\it both main phases} are parallelised. This is the main technical contribution of the paper. We clearly identify two main algorithmic stages: \textit{Tour construction} and \textit{Pheromone update}. A {\it data-parallelism} approach (which is theoretically better-suited to the GPU parallelism model than task-based parallelism) is described to enhance tour construction performance. Additionally, we describe various GPU design patterns  for the parallelisation of  the pheromone update, which has not been previously described in the literature.
 
The paper is organised as follows. We briefly introduce Ant Colony Optimisation for the TSP in Section \ref{sec:ant}, before describing
related work in Section \ref{sec:related}. In Section \ref{sec:GPUdesg} we present GPU designs for both main stages of the ACO algorithm. Experimental results are described in Section \ref{sec:hws}, before we conclude with a brief discussion and consideration of future work.

\section{Ant Colony Optimisation for the Travelling Salesman Problem}
\label{sec:ant}

The Travelling Salesman Problem (TSP) \cite{lawler1987traveling} involves finding the shortest (or ``cheapest") round-trip route that visits each of a number of ``cities" exactly once.  The symmetric TSP on $n$ cities may be represented as a complete weighted graph, $G$, with $n$ nodes, with each weighted edge, $e_{i,j}$, representing the inter-city distance $d_{i,j}=d_{j,i}$  between cities $i$ and $j$.
The TSP is a well-known {NP}-hard optimisation problem, and is used as a standard benchmark for many heuristic algorithms \cite{TSP}.

The TSP was the first problem solved by Ant Colony Optimisation (ACO) \cite{DORIGO2,DORIGO}. This method uses a number of simulated ``ants" (or {\it agents}), which perform distributed search on a graph. Each ant moves through on the graph until it completes a tour, and then offers this tour as its suggested solution. In order to do this, each ant may drop ``pheromone" on the edges contained in its proposed solution. The amount of pheromone dropped, if any, is determined by the {\it quality} of the ant's solution relative to those obtained by the other ants.  The ants probabilistically choose the next city to visit, based on {\it heuristic information} obtained from inter-city distances and the net pheromone trail. Although such heuristic information drives the ants towards an optimal solution, a process of ``evaporation" is also applied in order to prevent the process stalling in a local minimum.

The Ant System (AS) is an early variant of ACO, first proposed by Dorigo \cite{DORIGO}. The AS algorithm is divided into two main stages: \textit{Tour construction} and \textit{Pheromone update}. Tour construction is based on $m$ ants building tours in parallel. Initially, ants are randomly placed. At each construction step, each ant applies a probabilistic action choice rule, called the {\it random proportional rule}, in order to decide which city to visit next. The probability for ant $k$, placed at city $i$, of visiting city $j$ is given by the equation \ref{eq:prob}

\begin{equation}
p_{i,j}^{k}= \frac{\left[ \tau_{i,j}\right]^{\alpha} \left[ \eta_{i,j}\right]^{\beta}}{\sum_{l \in N_{i}^{k}}\left[ \tau_{i,l}\right]^{\alpha} \left[ \eta_{i,l}\right]^{\beta}}, \qquad if \;j \in N_{i}^{k},
\label{eq:prob}
\end{equation}

where $\eta_{i,j} = 1/d_{i,j}$ is a heuristic value that is available {\it a priori}, $\alpha$ and $\beta$ are two parameters which determine the relative {\it influences} of the pheromone trail and the heuristic information respectively, and $N_{i}^{k}$ is the feasible neighbourhood of ant $k$ when at city $i$. This latter set represents the set of cities that ant $k$ has not yet visited; the probability of choosing a city outside $N_{i}^{k}$ is zero (this prevents an ant returning to a city, which is not allowed in the TSP). By this probabilistic rule, the probability of choosing a particular edge $(i,j)$ increases with the value of the associated pheromone trail $\tau_{i,j}$ and of the heuristic information value $\eta_{i,j}$. Furthermore, each ant $k$ maintains a memory, $M^{k}$, called the {\it tabu list}, which contains the cities already visited, in the order they were visited. This memory is used to define the feasible neighbourhood, and also allows an ant to both to compute the length of the tour $T^{k}$ it generated, and to retrace the path to deposit pheromone. 

Another approach to tour construction is described in \cite{ACOBOOK}. This is based on exploiting the {\it nearest-neighbour} information of each city by creating a Nearest-Neighbour list of length $nn$ (between 15 and 40). In this case, an ant located in a city $i$ chooses the next city in a probabilistic manner among the $nn$ best neighbours. Once the ant has already visited all $nn$ cities, it selects the best neighbour according to the heuristic value given by the equation \ref{eq:prob}.

After all ants have constructed their tours, the pheromone trails are updated. This is achieved by first lowering the pheromone value on all edges by a constant factor, and then adding pheromone on edges that ants have crossed in their tours. Pheromone evaporation is implemented by

\begin{equation}
\tau_{i,j} \leftarrow (1-\rho)\tau_{i,j}, \qquad \forall(i,j) \in L,
 \label{equ:phero1}
\end{equation}

where $0 < \rho \leq  1$ is the pheromone evaporation rate. After evaporation, all ants deposit pheromone on their visited edges: 
\begin{equation}
\tau_{i,j} \leftarrow \tau_{i,j} + \sum_{k=1}^{m} \Delta \tau_{i,j}^{k},  
  \qquad \forall(i,j) \in L,
 \label{equ:phero2}
\end{equation}
 
where $\Delta \tau_{ij}$ is the amount of pheromone ant $k$ deposits. This is defined as follows:

\begin{equation}
 \Delta \tau_{i,j}^{k} = \left\{ % para la llave grandota
        \begin{tabular}{cc}
        	$1/C^{k}$ if\; $e(i,j)^{k}$ &  belongs to $T^{k}$ \\
        	$0$ &otherwise \\
        \end{tabular}
\right.
\label{equ:phero3}
\end{equation}

where $C^{k}$, the length of the tour $T^{k}$ built by the $k$-th ant, is computed as the sum of
the lengths of the edges belonging to $T^{k}$ . According to equation \ref{equ:phero3}, the better an ant's tour, the more pheromone the edges belonging to this tour receive. In general,
edges that are used by many ants (and which are part of short tours), receive more
pheromone, and are therefore more likely to be chosen by ants in future iterations of
the algorithm. 

\section{Related Work}
\label{sec:related}

St\"uzle \cite{STUZLE98} describes the simplest case of ACO parallelisation, in which independently instances of the ACO algorithm are run on different processors. Parallel runs have no communication overhead, and the final solution is taken as the best-solution over all independent executions.  Improvements over non-communicating parallel runs may be obtained by exchange information among processors. Michel and Middendorf \cite{Michel98} present a solution based on this principle, whereby separate colonies exchange pheromone information. In more recent work, Chen {\it et al.} \cite{Chen08} divide the ant {\it population} into equally-sized sub-colonies, each assigned to a different processor. Each sub-colony searches for an optimal local solution, and information is exchanged between processors periodically.  Lin {\it et al.} \cite{LIN}  propose dividing up the {\it problem} into subcomponents, with each subgraph assigned to a different processing unit. To explore a graph and find a complete solution, an ant moves from one processing unit to another, and messages are sent to update pheromone levels. The authors demonstrate that this approach reduces local complexity and memory requirements, thus improving overall efficiency. 

In terms of GPU-specific designs for the ACO algorithm, Jiening {\it et al.} \cite{ACOGPU1} propose an implementation of the Max-Min Ant System (one of many ACO variants) for the TSP, using C++ and NVIDIA Cg. They focus their attention on the tour construction stage, and compute the shortest path in the CPU. In \cite{ACOGPU2} You discusses a CUDA implementation of the Ant System for the TSP. The tour construction stage is identified as a CUDA kernel, being launched by as many threads as there are artificial ants in the simulation. 
The tabu list of each ant is stored in shared memory, and the pheromone and distances matrices are stored in texture memory. The pheromone update stage is  calculated on the CPU. You reports a 20x speed-up factor for benchmark instances up to 800 cities.
Li {\it et al.} \cite{ACOGPU3} propose a method based on a fine-grained model for GPU-acceleration, which maps a parallel ACO algorithm to the GPU through CUDA. Ants are assigned to single processors, and they are connected by a population-structure \cite{STUZLE98}.

Although these proposals offer a useful starting point when considering GPU-based parallelisation of ACO, they are deficient in two main regards. Firstly, they fail to offer any {\it systematic} analysis of how best to implement this particular algorithm. Secondly, they fail to consider an important component of the ACO algorithm; namely, the pheromone update. In the next Section we address both of these issues.

\section{GPU Designs for the ACO Algorithm}
\label{sec:GPUdesg}

In this Section we present several different GPU designs for the the Ant System, as applied to the TSP. The two main stages, \textit{Tour construction} and \textit{Pheromone update}, are deeply examined.  For tour construction, we begin by analysing traditional {\it task-based} implementations,  which motivates our approach of instead increasing the {\it data-parallelism}. For pheromone update, we describe several GPU techniques that are potentially useful in increasing application bandwidth.

\subsection{Tour construction kernel}

The ``traditional" {\it task-based} parallelism approach to tour construction is based on the observation that ants run in parallel looking for the best tour they can find. Therefore, any inherent parallelism exists at the level of individual ants. To implement this idea of parallelism on CUDA, each ant is identified as a CUDA thread, and threads are equally distributed among CUDA thread blocks.   Each thread deals with the task assigned to each ant; i.e, maintenance of an ant's memory (tabu list, list of all visited cities, and so on) and movement.

Using this na\"ive approach, each ant calculates the heuristic information to visit city $j$ from city $i$ according to equation \ref{eq:prob}. However, it is computationally expensive to repeatedly calculate those values for each computational step of each ant, $k$. Repeated computations of the heuristic information can be avoided by using an additional data structure, in which the heuristic values are stored, and are therefore calculated only once for each kernel call \cite{ACOBOOK}.  For the probabilistic choice of the next city by each ant, the tour construction kernel needs to generate random numbers on the GPU. 

The task-based parallelism just described presents several issues for GPU implementation. Fundamentally, it is not theoretically well-suited to GPU computation \cite{Tesla} \cite{Nickolls08}. This approach requires a relatively low number of threads on the GPU, since the recommended number of ants for solving the TSP problem is taken as the same as the number of cities \cite{ACOBOOK}. If, for example, 800 ants are needed to solve an 800 cities benchmark, the number of threads is too low to fully exploit the resources of the GPU. 

Moreover, the stochastic process required imply that the application presents an unpredictable memory access pattern. The final objection to this approach arises due to checking the list of cities visited; this operation presents many warp divergences (different threads in a warp take different paths), leading to serialisation \cite{Pguide}. 

\begin{figure}
 \centering
 \includegraphics[width=8.5cm,height=10.5cm]{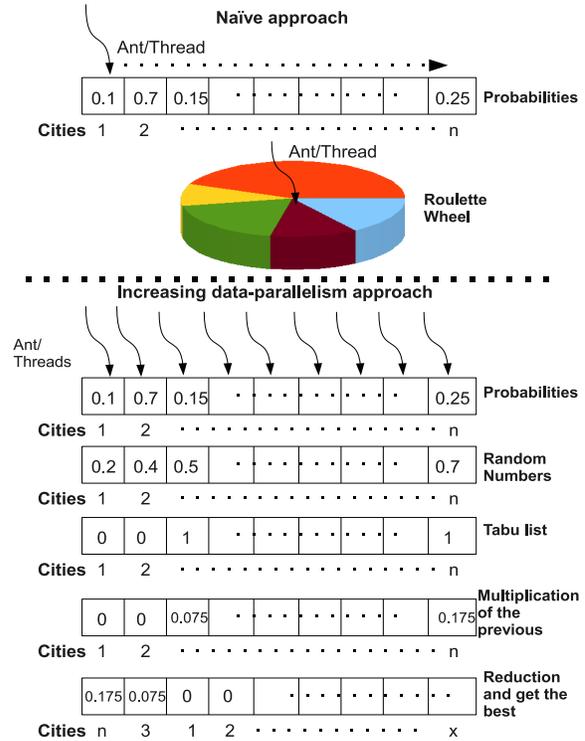}
 % SIMD.eps: 0x0 pixel, 0dpi, 0.00x0.00 cm, bb=
\vspace{-0.3cm}
 \caption{Increasing the SIMD parallelism on the tour construction kernel.}
 \label{fig:SIMD}
\end{figure}

Figure \ref{fig:SIMD} shows an alternative design, which increases the {\it data-parallelism} in the tour construction kernel, and also avoids warp divergences in the tabu list checking process. In this design, a {\it thread block} is is associated with each ant, and each thread in a thread block represents a city (or cities) the ant may visit. Thus, the parallelism is increased by a factor of $1:n$. 

A thread loads the heuristic value associated with its associated city, generates a random number in the interval $[0,1]$ to feed into the stochastic simulation, and checks whether the city has been visited or not. To avoid conditional statements (and, thus, warp divergences), the tabu list is represented as one integer value per each city, which can be placed in the register file (since it represents information private to each thread).  A city's value is 0 if it is visited, and 1 otherwise. Finally, these values are multiplied and stored in a shared memory array, which is then reduced to yield the next city to visit.

The number of threads per thread block on CUDA is a hardware limiting factor (see Table \ref{t:Tesla-features}). Thus, the cities should be distributed among threads to allow for a flexible implementation. A {\it tiling} technique is proposed to deal with this issue. Cities are divided into blocks (i.e. tiles). For each tile, a city is selected stochastically, from the set of unvisited cities on that tile. When this process has completed, we have a set of ``partial best" cities. Finally, the city with the best absolute heuristic value is selected from this partial best set.

The tabu list cannot be represented by a single integer register per thread in the tiling version, because, in that case, a thread represents more than one city. The 32-bit registers may be used on a bitwise basis for managing the list. The first city represented by each thread; i.e. on the first tile, is managed by bit 0 on the register that represents the tabu list, the second city is managed by bit 1, and so on. 

\subsection{Pheromone update kernel}
\label{sec:phero}

The last stage in the ACO algorithm is the pheromone update. This is implemented by
a kernel which comprise two main tasks: pheromone evaporation and pheromone deposit. 

\begin{figure}[t]
 \centering
 \includegraphics[scale=0.5]{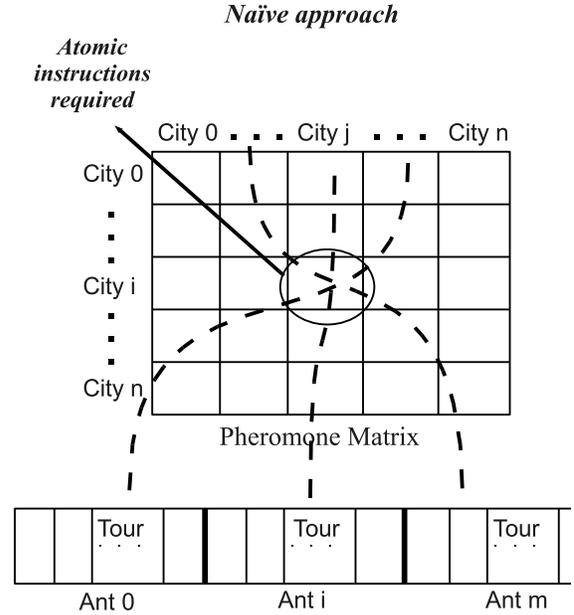}
 % Scatter.eps: 0x0 pixel, 0dpi, 0.00x0.00 cm, bb=
\vspace{-0.5cm}
 \caption{Pheromone Update kernel approach with atomic instructions.}
 \label{atomic}
\end{figure}

Figure \ref{atomic} shows the design of the pheromone kernel; this has a thread per city in an ant's tour. Each ant generates its own private tour in parallel, and they will feasibly visit the same edge as another ant. This fact forces us to use {\it atomic} instructions for accessing the pheromone matrix, which diminishes the application performance. Besides, those atomic operations are not supported by GPUs with CCC (CUDA Compute Capability) 1.x for floating point operations \cite{Pguide}.

\begin{figure}[t]
 \centering
 \includegraphics[scale=0.5]{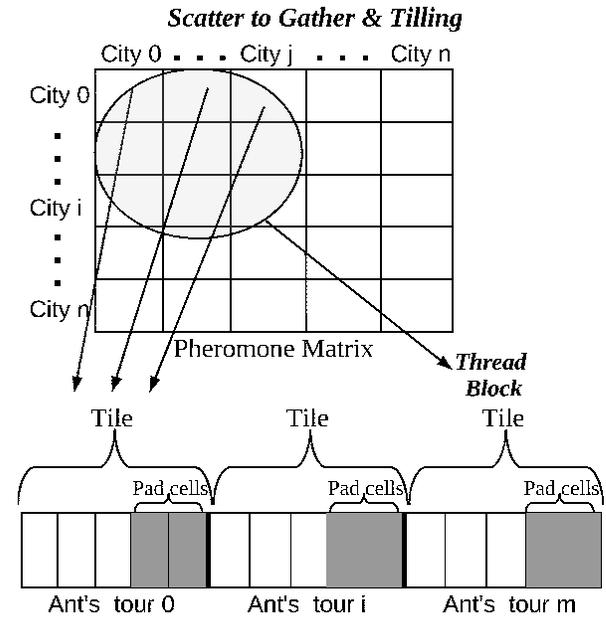}
 % Scatter.eps: 0x0 pixel, 0dpi, 0.00x0.00 cm, bb=
 \caption{Pheromone Update kernel approach by scatter to gather transformation.}
 \label{scatter}
\end{figure}

Therefore, a key objective is to avoid using atomic operations. An alternative approach is shown in Figure \ref{scatter}, where we use a {\it scatter} to gather transformations \cite{Scatter}. 

The configuration launch routine for the pheromone update kernel sets as many threads as there are cells are in the pheromone matrix ($c=n^{2}$) and equally distributes these threads among thread blocks. Thus, each cell is independently updated by each thread doing both the pheromone evaporation and the pheromone deposit. The pheromone evaporation is quite straightforward; we simply apply equation \ref{equ:phero1}. The pheromone update is a bit different. Each thread is now in charge of checking whether the cell represented by it has been visited by any ants; i.e. each thread accesses device memory to check that information.  This means that each thread performs $2*n^{2}$ memory loads/thread for a total of $l=2*n^{4}$ ($n^{2}$ threads) accesses of device memory. Notice that these accesses are 4 bytes each. Thus, the relation $loads:atomic$ is $l: c$. Therefore, this approach allows us to perform the computation {\it without} using atomic operations, but at the cost of drastically increasing the number of accesses to device memory.

A tiling technique is proposed for increasing the application bandwidth. Now, all threads cooperate to load data from global memory to shared memory, but they still access edges in the ant's tour. Each thread accesses global memory $2n^{2}/\theta$, $\theta$ being the tile size. The rest of the accesses are performed on shared memory. Therefore, the total number of global memory accesses is $\gamma=2n^{4}/\theta$. The relation $loads/atomics$ is lower $\gamma: c$, but maintains the orders of magnitude.  

We note that an ant's tour length (i.e. $n+1$) may be bigger than the maximum number of threads that each thread block can support (i.e. 512 threads/block for Tesla C1060). Our algorithm prevents this situation by setting our empirically demonstrated optimum thread block layout, and dividing the tour into tiles of this length. This raises up another issue; this is when $n+1$ is not divisible by the $\theta$. We solve this by applying padding in the ants tour array to avoid warp divergence (see Figure \ref{scatter}).

Unnecessary loads to device memory can be avoided by taking advantage of the problem's nature. We focus on the symmetric version of the TSP, so the number of threads can be reduced in half,  thus halving the number of device memory accesses. This so-called Reduction version actually reduces the overall number of accesses to either shared or device memory by having half the number of threads compared to the previous version. This is combined also with tiling, as previously explained.  The number of accesses per thread remains the same, giving a total of device memory access of $\rho=n_{4}/\theta$.

\section{Experimental Results}
\label{sec:hws}

We test our designs using a set of benchmark instances from the well-known TSPLIB library \cite{tsplib}
ACO parameters such as the number of ants $m$, $\alpha$, $\beta$, and so on are set according with the values recommended in \cite{ACOBOOK}. The most important parameter for the scope of this study is the number of ants, which is set $m=n$ (i.e., the number of cities).

We compare our implementations with the sequential code, written in ANSI C, provided by St\"uzle in \cite{ACOBOOK}. The performance figures are recorded for a single iteration, and averaged over 100 iterations. In this work we focus on the computational characteristics of the AS system and how it can be efficiently implemented on the GPU. The quality of the actual solutions obtained is not deeply studied, although the results are similar to those obtained by the sequential code for all our implementations.
 
\subsection{Performance evaluation}

The two main stages,  \textit{Tour construction} and \textit{Pheromone update}, are deeply evaluated on  two different GPU systems, both
based on the Nvidia Tesla. We use a C1060 model manufactured in mid 2008, and delivered as a graphics card plugged into a PCI-express 2 socket, and the more recent S2050 released in November 2010, and based on the Fermi architecture \cite{Fermi} (see Table ~\ref{t:Tesla-features} for full specifications).

\begin{table}[!h]
\caption{CUDA and hardware features for the Tesla C1060 GPU and the Tesla M2050.}\label{t:Tesla-features}
\begin{center}	

\begin{tabular}{|l|l|r|r|}
\hline 
GPU element    & Feature                   & Tesla C1060 &   Tesla M2050  \\ \hline 
Streaming      & Cores per SM  &       8     &        32      \\ % \hline
processors     & Number of SMs &      30     &        14      \\ % \hline 
(GPU           & Total SPs     &     240     &       448      \\ % \hline
cores)         & Clock frequency           &  1 296 MHz  &      1 147 MHz \\ \hline
Maximum        & Per multiprocessor        &   1 024     &    1 536       \\ % \hline 
number of      & Per block                 &     512     &    1 024       \\ % \hline 
threads        & Per warp                  &      32     &      32        \\ \hline
SRAM           & 32-bit registers          &   16 K~~    &     32 K~~     \\ % \hline
% SRAM & Constant (read-only) memory       &   64 KB     &     64 KB      \\ 
memory         & Shared memory             &   16 KB     & 16/48 KB \\ % \hline
available per  & L1 cache                  &      No     & 48/16 KB \\ % \hline
multiprocessor & (Shared + L1)  &   16 KB     &      64 KB     \\ \hline
               & Size                      &    4 GB     &      3 GB      \\ 
Global         & Speed                     &  2x800 MHz  &   2x1500 MHz   \\
(video)        & Width                     & 512 bits    &  384 bits      \\
memory         & Bandwidth                 & 102 GB/sc.  &  144 GB/sc.    \\
               & Technology                & GDDR3  &  GDDR5    \\ \hline

\end{tabular}

\end{center}
\end{table}

We first evaluate the existing, task-based approach, before assessing the impact of including various modifications.

\subsubsection{Evaluation of tour construction kernel}

\begin{table*}[htdp]
\caption{Execution times (in milliseconds) for various tour construction implementations (Tesla C1060).}\label{t:tgpus}
\begin{center}
\begin{footnotesize}
\begin{tabular}{|l|c|c|c|c|c|c|c|}
\hline
Code  version                           & \multicolumn{7}{c|}{TSPLIB benchmark instance (problem size)}  		\\ \cline{2-8}
                       & $att48$ & $kroC100$  & $a280$  & $pcb442$  & $d657$ & $pr1002$ & $pr2392$  		\\ \hline
1. Baseline Version        & 13.14 & 56.89 & 497.93 & 1201.52 & 2770.32 & 6181 & 63357.7 \\ \hline
2. Choice Kernel    &4.83 & 17.56 & 135.15 & 334.28 & 659.05 & 1912.59 & 18582.9 \\
3. Without CURAND &4.5&15.78&119.65&296.31&630.01&1624.05& 15514.9\\ 
4. NNList &2.36&6.39&33.08&72.79&143.36&338.88&2312.98\\
5. NNList + Shared Memory&1.81&4.42&21.42&44.26&84.15&203.15&2450.52 \\
6. NNList + Shared\&Texture Memory &1.35&3.51&16.97&38.39&75.07&178.3&2105.77\\
7. Increasing Data Parallelism &0.36&0.93&13.89&37.18&125.17&419.53&5525.76\\
8. Data Parallelism + Texture Memory & 0.34 & 0.91 & 12.12 & 36.57 & 123.17 & 417.72 & 5461.06\\ \hline
Total speed-up attained      & 38.09x &62.83x & 41.09x & 32.86x&22.49x & 14.8x&11.6x \\ \hline
\end{tabular}
\end{footnotesize}
\end{center}
\end{table*}

Table \ref{t:tgpus} summarises the evaluation results for different GPU strategies previously presented for the tour construction kernel. Our {\it baseline} version (1) is the na\"ive approach of {\it task-based} parallelism (that is, the approach that has been used to date). This redundantly calculates heuristic information. It is first enhanced by  (2) adding a kernel for avoiding redundant calculations; i.e. the Choice kernel. The increase in parallelism and the savings in terms  of operations drive this enhancement.  A slight enhancement (around 10-20\%) is obtained by (3) generating random numbers with a device function on the GPU, instead of using the NVIDIA CURAND library. Although randomness could, in principle, be compromised, this function is used by the sequential code. The next big enhancement in performance is obtained by (4) using the nearest-neighbour list (NNList). The NN List limits the generation of many costly random numbers. For a $NN=30$, we report up to 6.71x speed up factor for the biggest benchmark instance in the Table \ref{t:tgpus} ($pr2392$). Allocating the tabu list in the shared memory (5) enhances the performance for small-medium benchmark instances (up to 1.7x speed up factor). However, this trend is limited by the tabu list implementation being on a bitwise basis for biggest benchmarks. To manage this design, many modulo and integer divisions are required, which produces an extra overhead.  Using the texture memory (6) for random numbers gains a  25\% of performance improvement. Finally, our proposal of increasing the data-parallelism obtains the best speed up factor for the $att48$ benchmark, being close to 4x between 8 and 6 kernel versions. However, it tends to decrease along with the random number generation difference between both designs. Nevertheless, comparing both probabilistic designs ((3) and (8)), the speed up factor reaches up to 17.42x.

%We also analyse the thread block layout for configuring the kernel launch. It is worth noting the best thread block layout for the baseline %version and its derived optimisations is {\it not} the traditional one (128-256 threads/block). For the smallest benchmark, in which only few %threads are running at the same time, the best block size is 16 threads/block. This block size allows us to distribute more blocks among %SMs getting more parallelism, even though GPU resources were not fully occupied. Whenever the benchmark size increases, the best %block size configuration also increases, being the best thread block layout 64 threads/block. However, the data-parallelism approach finds %its best block size layout  for 128 threads/block for every benchmark.

\subsubsection{Evaluation of pheromone update kernel}

\begin{table*}[htdp]
\caption{Execution times (in milliseconds) for various pheromone update implementations (Tesla C1060).}\label{t:pgpus1}
\begin{center}
\begin{footnotesize}
\begin{tabular}{|l|c|c|c|c|c|c|}
\hline
Code version               & \multicolumn{6}{c|}{TSPLIB benchmark instance (problem size)}  		\\ \cline{2-7}
C1060                 & $att48$ & $kroC100$  & $a280$  & $pcb442$  & $d657$ & $pr1002$  		\\ \hline
1. Atomic Ins. + Shared Memory  & 0.15 & 0.35 & 1.76 & 3.45 & 7.44 & 17.45  \\ \hline
2. Atomic Ins.  & 0.16 & 0.36 & 1.99 & 3.74 & 7.74 & 18.23  \\
3. Instruction \& Thread Reduction & 1.18 & 3.8 & 103.77 & 496.44 & 2304.54 & 12345.4\\ 
4. Scatter to Gather + Tilling & 1.03 & 5.83 & 242.02 & 1489.88 & 7092.57 & 37499.2  \\
5. Scatter to Gather & 2.01 & 11.3 & 489.91 & 3022.85 & 14460.4 & 200201  \\ \hline

Total slow-down incurred   & 12.73x & 31.42x & 278.7x & 875.29x & 1944.23x & 11471.59x  \\ \hline
\end{tabular}
\end{footnotesize}
\end{center}
\end{table*}

\begin{table*}[htdp]
\caption{Execution times (in milliseconds) for various pheromone update implementations (Tesla M2050).} \label{t:pgpus2}
\begin{center}
\begin{footnotesize}
\begin{tabular}{|l|c|c|c|c|c|c|}
\hline
Code version                            & \multicolumn{6}{c|}{TSPLIB benchmark instance (problem size)}  		\\ \cline{2-7}
M2050                         & $att48$ & $kroC100$  & $a280$  & $pcb442$  & $d657$ & $pr1002$  		\\ \hline
1. Atomic Ins. + Shared Memory  & 0.04 & 0.09 & 0.43 & 0.79 & 1.85 & 4.22  \\ \hline
2. Atomic Ins.  & 0.04 & 0.09 & 0.45 & 0.88 & 1.98 & 4.37  \\
3. Instruction \& Thread Reduction & 0.83 & 2.76 & 88.25 & 501.32 & 2302.37 & 12449.9\\ 
4. Scatter to Gather + Tilling &0.8 & 4.45 & 219.8 & 1362.32 & 6316.75 & 33571  \\
5. Scatter to Gather & 0.66 & 4.5 & 264.38 & 1555.03 & 7537.1 & 40977.3  \\ \hline

Total slow-downs attained   &17.3x& 50.73 & 587.96x & 1737.95x & 3859.52x & 9478.68x \\ \hline
\end{tabular}
\end{footnotesize}
\end{center}
\end{table*}

In this case, the baseline version is our best-performing kernel version, which uses atomic instructions and shared memory.  From there, we show the slow-downs incurred by each technique. As previously explained, this kernel presents a tradeoff between the number of accesses to global memory for avoiding costly atomic operations and the number of those atomic operations (called $loads:atomic$). The ``scatter to gather" computation (5) pattern presents the major difference between both parameters. This imbalance is reflected in the performance degradation showed by the bottom-row on Table \ref{t:pgpus1}. The slow-down increases exponentially with the benchmark size, as expected. 

The tiling technique (4) improves the application bandwidth with the scatter to gather approach. The Reduction technique (3) actually reduces the overall number of accesses to either shared or device memory by having half the number of threads of versions 4 or 5. This also uses tiling to alleviate the pressure on device memory. Even though the number of loads per thread remains the same, the overall number of loads in the {\it application} is reduced. 

\begin{figure*}[th!]
\begin{center}
\subfigure[Nearest Neighbour List ($NN=30$). ]{
\label{fig:toura}
\includegraphics[scale=0.55]{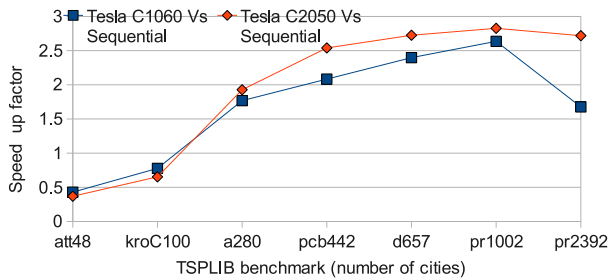}}
\subfigure[Fully probabilistic selection.]{
\label{fig:tourb}
\includegraphics[scale=0.55]{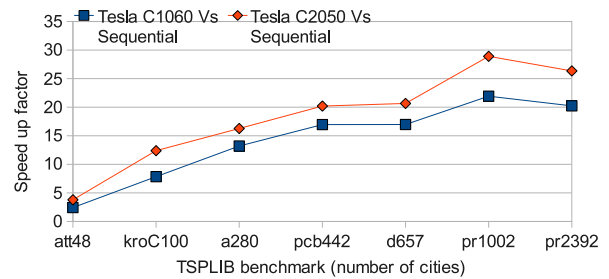}}

\caption{\label{fig:tour} Speed-up factor for tour construction kernel.}
\end{center}
\end{figure*}

\subsection{Overall performance}

Figure \ref{fig:tour} shows the speed-up factor between sequential code and both GPUs. Figure \ref{fig:toura} shows the speed-up obtained by simulating the Nearest Neighbour tour construction with 30 Nearest Neighbours ($NN=30$). This tour construction reduces the requirement for random number generation, and thus, the computation workload for the application. For the smallest benchmarks, the sequential code is faster than the GPU code. The number of ants, which is equivalent to the number of threads running in parallel on the GPU, is relatively small for these instances; i.e. 48, 100. Besides, those threads are quite heavy-weight threads based on task-parallelism.   The CPU is not only theoretically, but now {\it empirically demonstrated} to be, better suited to deal with this coarse-grained task. 

However, the GPU obtains better performance as long as the benchmark size increases, reaching up to 2.65 x on Tesla C1060 and 3x on Tesla C2050. We note that the maximum performance is obtained for the pr1002 benchmark, after which point the performance improvement begins to decrease. This behaviour is even worse for the Tesla C1060. From that benchmark, the GPU occupancy is drastically affected, and for the Tesla C1060 the tabu list can only be located on a bit bases in shared memory, which introduces an extra overhead. 

Figure \ref{fig:tourb} shows the effect of implementing our proposal for increasing the data-parallelism, compared to the fully probabilistic version of the sequential code. We observe an up to 22x speed up factor for Tesla C1060, and up to 29x for Tesla M2050. 
This version presents much more fine-grained parallelism, where the threads are light-weight. This is reflected in the performance enhancement on the smallest benchmarks. However, this version launches as many random numbers as number of cities, and performs a reduction each tiling step. This begins to negatively affect performance for biggest the biggest benchmark instances, such as pr2392.  

\begin{figure}
 \centering
 \includegraphics[scale=0.6]{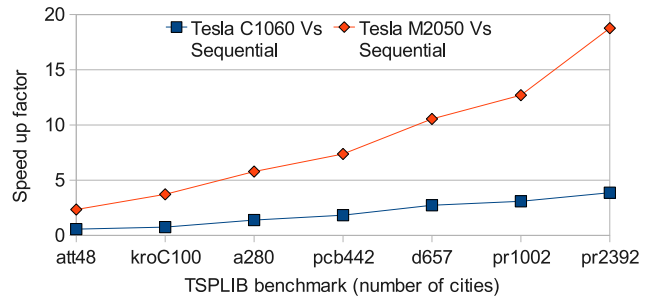}
 % PheromoneUpdate.eps: 0x0 pixel, 0dpi, 0.00x0.00 cm, bb=
 \caption{Speed-up factor for pheromone update kernel.}
 \label{fig:pheroeval}
\end{figure}

Figure \ref{fig:pheroeval} shows the speed-up factor for the best version of the pheromone update kernel compared to the sequential code. The pattern of computation for this kernel is based on data-parallelism, showing a linear speed-up along with the problem size. However, the lack of supporting atomic operations on Tesla C1060 for floating points operations means that, for the smallest benchmark instances, the sequential code obtains better performance. As long as the level of parallelism increases, the performance also increases, obtaining up to 3.87x speed-up for Tesla C1060 and 18.77x for Tesla M2050.
 
\section{Conclusions and Future Work}

Ant Colony Optimisation (ACO) belongs to the family of population-based meta-heuristic that has been successfully applied to many NP-complete problems. As a population-based algorithm, it is intrinsically parallel, and thus well-suited to implementation on parallel architectures.  The ACO algorithm comprises two main stages; \textit{tour construction} and \textit{pheromone Update}.  Previous efforts for parallelizing ACO on the GPU focused on the former stage, using task-based parallelism. We demonstrated that this approach does not fit well on the GPU architecture, and provided an alternative approach based on {\it data parallelism}. This enhances the GPU performance by both increasing the parallelism and avoiding warp divergence.

In addition, we provided the first known implementation of the \textit{pheromone update} stage on the GPU. In addition, some GPU computing techniques were discussed in order to avoid atomic instructions. However, we showed that those techniques are even more costly than applying atomic operations directly.

Possible future directions will include investigating the effectiveness of GPU-based ACO algorithms on other NP-complete optimisation problems. We will also implement other ACO algorithms, such as the Ant Colony System, which can also be efficiently implemented on the GPU. The conjunction of ACO and GPU is still at a relatively early stage; we emphasise that we have only so far tested a relatively simple variant of the algorithm. There are many other types of ACO algorithm still to explore, and as such, it is a potentially fruitful area of research. We hope that this paper stimulates further discussion and work.
 
\section*{Acknowledgement}

This work was partially supported by a travel grant from HiPEAC, the European Network of Excellence on High Performance and Embedded Architecture and Compilation (http://www.hipeac.net).

\bibliographystyle{IEEEtran}
\bibliography{ACO}

% \end{thebibliography}

\end{document}